\documentclass[12pt]{article}
\usepackage{amssymb}
\usepackage{latexsym}
\usepackage{amscd}

\newtheorem{tom}{Theorem}
\newtheorem{tom1}{Theorem}

\newtheorem{lem}{Lemma}[subsection]

\newtheorem{dfn}{Definition}[subsection]
\newtheorem{prop}{Proposition}[subsection]

\begin{document}

\thispagestyle{empty}

\begin{flushright}
\vspace*{1mm}
\hfill{hep-th/0212307}\\
\hfill{math.DG/0212307}\\
\hfill{SU-ITP-02/49}
\end{flushright}

\vskip 2cm

\begin{center}

{\large \bf GEOMETRIC MODEL FOR COMPLEX}

\medskip
{\large \bf NON-K\"AHLER MANIFOLDS WITH $SU(3)$}

\medskip
{\large \bf STRUCTURE}

\vskip 1.5cm

{\bf Edward Goldstein$^1$ and Sergey Prokushkin$^2$}

\vskip 1cm

{$^1$\it Department of Mathematics, Stanford University,}
{\it Stanford, CA 94305, USA}\\
{\tt egold@math.stanford.edu}

\medskip
{$^2$\it Department of Physics, Stanford University,}
{\it Stanford, CA 94305, USA}\\
{\tt prok@itp.stanford.edu}

\end{center}

\vskip 2cm

\begin{abstract}
For a given complex $n$-fold $M$ we present an explicit construction of all complex $(n+1)$-folds which are principal holomorphic $T^2$-fibrations over $M$. For physical applications we consider the case of $M$ being a Calabi-Yau $2$-fold. We show that for such $M$, there is a subclass
of the $3$-folds that we construct, which has natural families of non-K\"ahler $SU(3)$-structures satisfying 
the conditions for $\mathcal{N} = 1$ supersymmetry in the heterotic string theory compactified on the $3$-folds.  
We present examples in the aforementioned subclass with $M$
being a $K3$-surface and a $4$-torus.
   
\end{abstract}

\newpage

\section{Introduction}

Recently, 6-dimensional non-K\"ahler manifolds with $SU(3)$ structure have attracted considerable attention in physics literature as possible internal spaces for supersymmetric string compactifications \cite{bd, kstt, louis, lust, bbdg, gm, tt}.
In particular, such manifolds appear in the T-dual description of the type IIB theory compactifications
on a Calabi-Yau space in the presence of background RR and NS fluxes (see e.g. \cite{kstt, louis} for detailed discussion
and further references), and in the heterotic string theory compactifications with non-zero background torsion $H$
(see e.g. \cite{Str, bd, lust}).  Non-zero background fluxes induce a low-energy potential that stabilizes many
of the geometric moduli, and therefore play an important role in superstring
compactifications (see e.g. \cite{kst} and references therein). In many different contexts, 
the ``manifolds with intrinsic torsion'' or ``manifolds with 
$SU(3)$ structure'' have been studied for a long time (see e.g. \cite{SS, Hull, FFS, KM, Vafa, joyce, fi, salamon1, drs,
salamon, CS, rocek, ip, gkmw, kmpt, hmw}).

Remarkably, all these manifolds are some $T^2$ bundles over Calabi-Yau varieties, 
see \cite{bd, kstt, louis, lust, bbdg}. In most of known examples the metric has the following 
local form: 
\begin{equation}
\label{metric}
                 g_{base} + (dx+\alpha)^2+(dy+\beta)^2  \,,
\end{equation}
where $g_{base}$ is the metric on the base (usually Calabi-Yau), $x$ and $y$ are local coordinates such that 
$dx+i\, dy$ is a holomorphic form on the $T^2$-fibers, and $\alpha$ and $\beta$ are (local) 1-forms on the base.

One can ask the question of how to construct such $T^2$-fibrations over various base Calabi-Yau $n$-folds, and what conditions should $\alpha$ and $\beta$ satisfy for such a complex $T^2$-bundle to exist over a Calabi-Yau manifold. 
In this paper we produce a general construction of families $M^{P,Q}$ 
of holomorphic $T^2$-fibrations over a base Hermitian $n$-fold $(M,g)$ (A Hermittian $n$-fold $(M,g)$ is a complex manifold $M$ of complex dimension $n$ with a Hermititan metric $g$ on it). Our first result is the following theorem:
\begin{tom}
\label{main}
Let $\omega_P$ and $\omega_Q$ be closed 2-forms on a complex Hermitian $n$-fold $(M,g)$ s.t. the following two conditions hold:\\
1) $\omega_P+i\omega_Q$ has no component in $\Lambda^{0,2}T^{\ast}M$.\\
2) $\frac{\omega_P}{2\pi}$ and $\frac{\omega_Q}{2\pi}$ represent integral cohomology classes. \\
Then there is a complex, Hermitian $(n+1)$-fold $(M^{P,Q},g')$ with a free structure-preserving $T^2$-action and a holomorphic fibration 
$\pi:M^{P,Q} \mapsto M^{P,Q}/T^2 \simeq M $ such that the following holds:\\
A) For any 1-forms $\alpha$ and $\beta$ defined on some open subset of $M$ and satisfying $d\alpha=\omega_P$ and $d\beta=\omega_Q$ there are local coordinates $x$ and $y$ on $M^{P,Q}$ such that $dx+i\, dy$ is a holomorphic form on the $T^2$-fibers and the metric on $M^{P,Q}$ has the form 
 (\ref{metric}) with $g_{base}$ being $g$.\\
B) If $M$ admits a non-vanishing holomorphic $(n,0)$-form $\Omega^{n,0}$ then  $M^{P,Q}$ admits a non-vanishing holomorphic $(n+1,0)$-form $\Omega^{n+1,0}$. \\
C) If either $\omega_P$ or $\omega_Q$ represent a non-trivial cohomology class 
then $M^{P,Q}$ admits {\bf NO} K\"ahler metric.
\end{tom}
We believe that Theorem~\ref{main} generalizes all known examples of complex non-K\"ahler 
manifolds so far used as internal manifolds in supersymmetric compactifications both in heterotic and type II theories.\\
Note that in the theorem above, $M$ does not have to be a Calabi-Yau, it is just a complex manifold with a Hermitian metric. Thus our construction can also be viewed as a new construction of complex manifolds which includes as a special case Hopf manifolds (for those $\omega_Q=0$ in our terminology) and the complex structure (found by E. Calabi in \cite{Cal}) on the product of odd-dimensional spheres $S^{2m+1} \times S^{2k+1}$.\\
Also the manifolds $M^{P,Q}$ that we construct, are principal $T^2$-bundles for the standard complex structure on $T^2$ and the Hermitian metric on $M^{P,Q}$ is $T^2$-invariant. Vice versa we have the following theorem:

\begin{tom}
\label{class}
Let $(N,g)$ be a Hermitian $(n+1)$-fold with a free, structure-preserving $T^2$-action. Assume also that $g$ restricted to the $T^2$-fibers is the standard metric on $T^2$. Then $N$ is isomorphic (biholomorphic and isometric) to a manifold $M^{P,Q}$ for some Hermitian $n$-fold $(M,g')$ and the closed $2$-forms $\omega_P$ and $\omega_Q$ on $M$ satisfying the conditions of Theorem \ref{main}.
\end{tom}
We remark that if $N$ is a complex manifold with a free holomorphic $T^2$-action when we can find a Hermitian metric on $M$ satisfying the conditions of Theorem \ref{class}, thus $N$ is biholomorphic to a manifold $M^{P,Q}$ we construct.
Thus many other known examples of principal $T^2$-fibrations (e.g. Iwasawa manifolds) fall under our construction.

Note also that we can relax the condition 1) of the theorem and construct a subclass of  
{\bf non-complex} {\it half-flat} manifolds considered in \cite{louis, lust, gm}. These manifolds will be relevant 
in the models of string compactification in which the superpotential stabilizes the almost complex structure
to a value that is not integrable. 

Next we will specialize to the case with $n=2$. From now on we consider $(M,g_{CY},\Omega^{2,0})$ to be a Calabi-Yau $2$-fold with a non-vanishing holomorphic $(2,0)$-form $\Omega^{2,0}$ which has unit length with respect to $g_{CY}$. If we pick the forms $\omega_P$ and $\omega_Q$ on $M$ satisfying the conditions of Theorem \ref{main} then we obtain a Hermitian $3$-fold $M^{P,Q}$ with a holomorphic $(3,0)$-form $\Omega^{3,0}$ of unit length. To fix notation we make the following definition:
\begin{dfn}
\label{special}
An $SU(3)$-structure on an almost-complex $3$-fold $N$ is a Hermitian metric $g$ on $N$ and a $(3,0)$-form $\Omega^{3,0}$ (not necessarily holomorphic) of unit length with respect to $g$. \\
Suppose $(N,g)$ is a complex (integrable) $3$-fold with a Hermitian metric $g$ and suppose there is a non-vanishing holomorphic $(3,0)$-form $\Omega^{3,0}$ on $N$. Then $(N,g,\frac{\Omega}{|\Omega|})$ is called a {\it special} $SU(3)$-structure on $N$. The function $\phi=\frac{1}{8}log|\Omega|: N \mapsto \mathbb{R}$ is called the dilaton.
\end{dfn}
In \cite{Str},  the  ${\mathcal N}=1$ supersymmetry conditions for heterotic string compactifications 
have been studied (see also \cite{lust}). In these models, the internal six-dimensional manifold $N$ must have 
a special $SU(3)$-structure $(N,g,\frac{\Omega}{|\Omega|})$. 
The supersymmetry condition on the internal manifold is:
\begin{equation}
\label{strom}
\ast \ d \ast \tilde{\omega}=i(\overline{\partial}-\partial)log |\Omega^{3,0}|
\end{equation}
Here $\tilde{\omega}$ is the Hermitian $(1,1)$-form on $N$ - it is often denoted by $J$ in the physics literature. This condition has the following geometric significance: On every Hermitian $n$-fold $N$ there is a unique connection $\nabla^B$ (the Bismut connection, \cite{Gaud, fg}) which preserves both the metric and the complex structure and those torsion $H$ is a skew-symmetric $3$-form. Equation (\ref{strom}) implies that the holonomy of this connection is contained in $SU(3)$ (\cite{Str}).

The equation (\ref{strom}) was rewritten in \cite{lust} in terms of the intrinsic torision of the $SU(3)$-structures. The intrinsic torsion of the (gereral) $SU(3)$ structures falls into five different classes
$W_1\oplus W_2\oplus W_3\oplus W_4\oplus W_5$ (see \cite{salamon, lust}).
In this classification,
our manifolds $M^{P,Q}$ have $W_1=0$, $W_2=0$, which is another way to see that the almost complex
structure is integrable. In our main construction with the metric of a form (\ref{metric}), we have also 
$W_5=0$,  so that the holomorphic $(3,0)$-form $\Omega^{3,0}$ has constant norm. In the language of the heterotic
string theory, that means that the dilaton field is constant and there is no warp factor (see \cite{Str}).
For a $3$-fold $N$ with a special $SU(3)$-structure as in Definition \ref{special}, the requirement (\ref{strom}) of ${\mathcal N}=1$ supersymmetry in the heterotic theory is {\bf equivalent} to the following ``torsional constraints'' \cite{salamon, lust}
\begin{equation}
\label{torscon}
      2W_4 + W_5 = 0 
\end{equation}
and
\begin{equation}
\label{exact}
W_4 ~ and ~ W_5 ~ are ~ ~ real ~ ~ and ~ ~ exact ~
\end{equation}
In our class $M^{P,Q}$ this leaves the manifolds with only non-zero torsion in the class $W_3$, so-called {\it special Hermitian} manifolds. 
These are complex {\it half-flat} manifolds \cite{salamon}, and the Hitchin flow of those manifolds (see \cite{Hitchin, lust}) produces manifolds with $G_2$-holonomy. We prove the following
\begin{tom}
\label{cd}
Let $(M,g_{CY},\Omega^{2,0})$ be a Calabi-Yau $2$-fold with a non-vanishing holomorphic $(2,0)$-form $\Omega^{2,0}$. Let $\omega_P$ and $\omega_Q$ be closed $2$-forms on $M$ satisfying the conditions of Theorem \ref{main}. Then the $3$-fold $M^{P,Q}$ has a natural special $SU(3)$-structure with a {\bf constant} dilaton and moreover the torsional constraints (\ref{torscon}),(\ref{exact}) hold if and only if the $(1,1)$-components of $\omega_P$ and $\omega_Q$ are anti-selfdual: 
$\star\omega^{(1,1)}_P = -\omega^{(1,1)}_P$,
$\star\omega^{(1,1)}_Q = -\omega^{(1,1)}_Q$.
\end{tom}
We continue to consider the manifolds in the situation of Theorem \ref{cd}. 
On such manifolds, we consider a modified construction with the Hermitian metric of a form
\begin{equation}
\label{metricwarp}
         g_{\psi} =     e^{2\psi} g_{CY} + (dx+\alpha)^2+(dy+\beta)^2
\end{equation}
where $\psi$ in the warp factor in front of the Calabi-Yau metric $g_{CY}$, is an arbitrary function of the 
base coordinates. This function accounts for a non-constant dilaton background which is typical for  
the heterotic string compactifications. Now the holomorphic $(3,0)$-form $\Omega^{3,0}$ on $M^{P,Q}$ 
has non-constant norm. We prove the following theorem:
\begin{tom}
\label{dilaton}
Let $(M,\omega_{CY},\Omega^{2,0})$ be a Calabi-Yau $2$-fold with a non-vanishing holomorphic $(2,0)$-form $\Omega^{2,0}$. Let $\omega_P$ and $\omega_Q$ be closed $2$-forms on $M$ satisfying the conditions of Theorem \ref{main}. Assume also that the $(1,1)$-components of $\omega_P$ and $\omega_Q$ are anti-selfdual. Let $\psi: M \mapsto \mathbb{R}$ be any function. Then there is a special $SU(3)$-structure on $M^{P,Q}$ with the Hermitian metric $g_{\psi}$ of the form (\ref{metricwarp}) where $\alpha$ and $\beta$ are as Theorem \ref{main} and the torsional constraints (\ref{torscon}),(\ref{exact}) hold.
\end{tom}
For $M$ compact, our construction satisfies the topological 
constraints pointed out in \cite{Str}: $h^{3,0} = 1$, $c_1 = 0$. Also, our construction has zero Euler characteristic. 
Next we will present examples where Theorem \ref{dilaton} holds for $M$  being a $K3$-surface and a $4$-torus.

One comment is now in order. The condition 2) in the Theorem~\ref{main}, which arose here as one of 
the sufficient conditions for the geometric construction to exist, can be motivated by physical arguments.
Namely, it is equivalent to 
the Dirac quantization condition of the NS flux in string theory. Indeed, a model with the metric (\ref{metric}) 
can be considered as T-dual (after two T-dualities along $T^2$ fibers) to a type IIB model with the 
internal 6-manifold $\tilde{N} \simeq CY \times T^2$ (see, e.g. the discussion in \cite{kstt}) and the background 
NS $\mathcal{B}$ field 
\begin{equation}
\label{B}
             \mathcal{B} = -dx \wedge \alpha - dy \wedge \beta
\end{equation}
(compare to formulas (4.23) and (4,24) of \cite{kstt}), or background NS flux $\mathcal{H}_3$
\begin{equation}
\label{H}
             \mathcal{H}_3 = d\mathcal{B} =  dx \wedge \omega_P + dy \wedge \omega_Q  \,,
\end{equation}
with $d\alpha=\omega_P$ and $d\beta=\omega_Q$ defined on the base $CY$.
The flux $\mathcal{H}_3$ satisfies the Dirac quantization condition (see eq.~(2.1) in \cite{kstt}):
\begin{equation}
\label{quant}
        {1\over {(2\pi)^2}} \int_\gamma \mathcal{H}_3 
         = n_{\gamma} \in \mathbb{Z} \,, \quad \gamma \in H_3(\tilde{N}, \mathbb{Z})     \,, 
\end{equation}
from which it follows that 
\begin{equation}
\label{quant2}
        \int_{\gamma_{CY}}  {\omega_P\over {2\pi}}
         = n_{\gamma} \in \mathbb{Z} \quad 
        \int_{\gamma_{CY}}  {\omega_Q\over {2\pi}}
         = m_{\gamma} \in \mathbb{Z}\,,  \quad \gamma_{CY} \in H_2(CY, \mathbb{Z})        \,, 
\end{equation}
i.e., condition 2).\\

We will also show that one can lift Special Lagrangian submanifolds and
fibrations from the Calabi-Yaus to the bundles $M^{P,Q}$ in the case $\omega_Q=0$, and Special 
Lagrangian submanifolds on those
bundles will still be calibrated by $Re\;  \Omega^{3,0}$ (and hence minimal).

\section{Global model -- the construction}\label{solut}

In this section we prove the following theorem:
\begin{tom1}
Let $\omega_P$ and $\omega_Q$ be closed 2-forms on a complex Hermitian $n$-fold $(M,g)$ s.t. the following two conditions hold:\\
1) $\omega_P+i\omega_Q$ has no component in $\Lambda^{0,2}T^{\ast}M$.\\
2) $\frac{\omega_P}{2\pi}$ and $\frac{\omega_Q}{2\pi}$ represent integral cohomology classes. \\
Then there is a complex, Hermitian $(n+1)$-fold $(M^{P,Q},g')$ with a free structure-preserving $T^2$-action and a holomorphic fibration 
$\pi:M^{P,Q} \mapsto M^{P,Q}/T^2 \simeq M $ such that the following holds:\\
A) For any 1-forms $\alpha$ and $\beta$ defined on some open subset of $M$ and satisfying $d\alpha=\omega_P$ and $d\beta=\omega_Q$ there are local coordinates $x$ and $y$ on $M^{P,Q}$ such that $dx+i\, dy$ is a holomorphic form on the $T^2$-fibers and the metric on $M^{P,Q}$ has the form 
 (\ref{metric}) with $g_{base}$ being $g$.\\
B) If $M$ admits a non-vanishing holomorphic $(n,0)$-form $\Omega^{n,0}$ then  $M^{P,Q}$ admits a non-vanishing holomorphic $(n+1,0)$-form $\Omega^{n+1,0}$. \\
C) If either $\omega_P$ or $\omega_Q$ represent a non-trivial cohomology class 
then $M^{P,Q}$ admits {\bf NO} K\"ahler metric.
\end{tom1}
{\bf Proof:} First pick two line bundles $P$ and $Q$ s.t. the first Chern class $c_1(P)=[- \frac{\omega_P}{2\pi}]$ and $c_1(Q)=[ -\frac{\omega_Q}{2\pi}]$. This is certainly possible since there is a 1-1 correspondence between smooth line bundles and $H^2(M ,\mathbb{Z})$. Put some Hermitian metrics on $P$ and $Q$. We have the following elementary
\begin{lem}
One can choose a Riemannian connection $\nabla$ on $P$ whose curvature form is $\omega_P$. The analogous statement is true for $Q$.
\end{lem}
{\bf Proof:} Let $\nabla'$ be some Riemannian connection on $P$ and let $\omega$ be its curvature. Then $\omega_P-\omega$ is trivial cohomologically, hence $\omega_P-\omega=d\alpha$ for some 1-form $\alpha$. Now the connection $\nabla=\nabla'+i\alpha$ is a Riemannian connection on $P$ whose curvature form is $\omega_P$.   Q.E.D.\\
We choose Riemannian connections on $P$ and $Q$ as in the previous Lemma and proceed as follows:  Consider the total space of the direct sum $P \oplus Q$ over $M$. The connections $\nabla$ and $\nabla'$ give rise
to a connection  on $P \oplus Q$ and a horizontal distribution $H$ that
is a subspace of $T(P\oplus Q)$. The horizontal distribution is obtained as follows: given a curve $\gamma(t)$ in $M$ and a pair $(\xi,\eta)$ in the fiber of $P \oplus Q$ over $\gamma(0)$, we have a unique curve $\gamma_H(t)=(\xi(t),\eta(t))$ in $P \oplus Q$ over $\gamma(t)$  such that $\xi(t)$ and $\eta(t)$ are parallel along $\gamma$. The tangent vector $\gamma_H'(t) $ lives in the horizontal distribution $H$ and it is called the {\it Horizontal Lift} of the tangent vector $\gamma'(t)$. \\
For each point $p \in M$ let $S_1(p)$
be the
unit circle bundle of $P$ over $p$ and let $S_2(p)$ be the unit circle
bundle of
$Q$ over $p$. Let $T(p)=S_1(p) \times S_2(p)$ and let \[M^{P,Q} =\bigcup
T(p)\]
Thus $M^{P,Q}$ is a 2-torus bundle over $M$. The distribution $H$ along
$M^{P,Q}$ is
 tangent to $M^{P,Q}$: this is because the parallel transport preserves length.\\
To understand $H$ along $M^{P,Q}$ better let $\xi$ be a local unit length section of $P$ on $M$ and $\eta$ be a
local unit length section of $Q$ on $M$ . The sections $\xi,\eta$ define local
coordinates $ x,y$ on $M^{P,Q}$, namely any point $z \in M^{P,Q}$ can be written as
$(e^{ix} \xi,e^{iy}\eta)$. Also 
$\xi$ defines a connection 1-form $\alpha'$ on $M$ by
\[\nabla \xi = \alpha' \otimes \xi\] 
This means that for any tangent vector $v$ to $M$ we have $\nabla_v \xi=\alpha'(v)\xi$. Now $\alpha'$ is imaginary valued and 
\begin{equation}
\label{curvature}
\omega_P=-i\,d\alpha'
\end{equation}
Similarly $\eta$ defines a connection 
1-form $\beta'$
on $M$ by \[\nabla \eta = \beta' \otimes \eta\] The forms $\alpha'$ and
$\beta'$ are purely imaginary and the horizontal space $H$ is precisely the 
kernel of the two 1-forms
\begin{equation}
\label{im}
           i\,dx+\pi^{\ast}\alpha' ~ and ~ i\, dy+\pi^{\ast}\beta'
\end{equation}
Indeed let $(\xi_H(t),\eta_H(t))$ be a curve in $M^{P,Q}$ sitting over a curve $\gamma(t)$ in $M$ such that $\xi_H(t)$ and $\eta_H(t)$ are parallel. We can write \[(\xi_H(t),\eta_H(t))=(e^{ix(t)}\xi(\gamma(t)),e^{iy(t)}\eta(\gamma(t)))\]
The condition that $\xi_H$ is parallel is equivalent to 
\[0 = ix'(t)e^{ix}\xi+e^{ix}\nabla_{\gamma'}\xi=e^{ix}(i\,dx(\gamma_H')+\pi^{\ast}\alpha'(\gamma_H'))\xi\]
which is equivalent to saying that $\gamma_H'$ is in the kernel of $i\,dx+\pi^{\ast}\alpha'$. Similarly the fact that $\eta_H$ is parallel is equivalent to saying that $\gamma_H'$ is in the kernel of $i\, dy+\pi^{\ast}\beta'$.\\ 
Let $V$ be the vertical space of $M^{P,Q}$- the
tangent space to the fibers. On every fiber $T(p)$ we have a natural 
$S^1 \times
S^1=T^2$-action given by \[(e^{ix},e^{iy}) \cdot(\xi,\eta)=
(e^{ix}\xi,e^{iy}\eta)\] We
have vector fields $\partial_x$ and $\partial_y$ tangent to the fibers. 
We define the complex structure on $T(p)$ to be the
natural one: $\partial_x \mapsto \partial_y$ and $\partial_y 
\mapsto -\partial_x$.
The almost complex structure on $H$ is induced from the projection onto $M$. Thus
$M^{P,Q}$ acquires an almost complex structure $I$. 
Define now a $(1,0)$-form $\rho$ on $M^{P,Q}$ by requiring that
\begin{equation}
\label{}
     \rho=0  ~ on ~H ~ and ~\rho=dx+i\, dy ~ on ~V
\end{equation}
From equation (\ref{im}) we conclude that 
\begin{equation}
\label{rho}
\rho=(dx-i\alpha')+i(dy-i\beta')
\end{equation}
Also pick a local holomorphic $(n,0)$-form $\Omega^{n,0}$ on $M$ and define an $(n+1,0)$-form $\Omega^{n+1,0}$ on $M^{P,Q}$ by
\begin{equation}
\label{varphi2}
\Omega^{n+1,0}= \rho \wedge \pi^{\ast}(\Omega^{n,0})
\end{equation}
We compute from (\ref{curvature}) that 
\begin{equation}
\label{dvarphi2}
d\Omega^{n+1,0}=\pi^{\ast}((\omega_P+i\omega_Q) \wedge \Omega^{n,0})=0
\end{equation}
Note that if $(\omega_P+i\omega_Q)$ had a non-zero component of type $(0,2)$ then $d\Omega^{n+1,0}$ 
would have had a non-zero $(n,2)$-component and the almost complex structure would not be integrable. 
But in our case $d\Omega^{n+1,0}=0$ and so the almost complex structure is integrable. This implication is
standard: by Newlander-Nirenberg theorem it is enough to prove that for a
$(1,0)$-form
$\theta$ we have that $d\theta$ is of type $\Lambda^{2,0} \oplus
\Lambda^{1,1}$.
 Now we have $0= \theta \wedge \Omega^{3,0}$. Taking the exterior derivative
we get
\[0= d\theta \wedge \Omega^{3,0}\] i.e.  $d\theta$ is of type $\Lambda^{2,0}
\oplus \Lambda^{1,1}$.\\
Also if $\Omega^{n,0}$ is a holomorphic, non-vanishing $(n,0)$-form defined on the whole of $M$ then $\Omega^{n+1,0}$ is a holomorphic, non-vanishing $(n+1,0)$-form on $M^{P,Q}$ and this proves {\bf B)}.\\
Both $H$ and $V$  have a natural
Hermitian metric and thus $M^{P,Q}$ is naturally a Hermitian manifold. Let $\alpha$ and $\beta$ be any 1-forms on $M$ s.t. $d\alpha=\omega_P$ and $d\beta=\omega_Q$. Then we can find local unit length sections $\xi$ of $P$ and $\eta$ of $Q$ s.t. $i\alpha$ and $i\beta$ are the connection 1-forms defined by $\xi$ and $\eta$, see (\ref{curvature}) . Now $\xi$ and $\eta$ define local coordinates $x$ and $y$ on $M^{P,Q}$ as before and the metric on $M^{P,Q}$ has the form as in equation (\ref{metric}).\\
We are still left to prove that if either $\omega_P$ or $\omega_Q$ is non-trivial in cohomology then $M^{P,Q}$ admits no K\"ahler metric.  One can also easily show that the fibers of $\pi$ are $0$ in the real 2-dimensional homology of $M^{P,Q}$,
see more details about homology in the Section 7 of this paper. The triviality of the (complex) fibers in homology implies that there is no K\"ahler form on $M^{P,Q}$: for such a form would integrate to a positive number on the fibers.  Q.E.D.

\section{Holomorphic principal T2 fiber bundles}

The goal of this section is to state and prove Theorem 2 from the Introduction: 
\begin{tom1}
Let $(N,g)$ be a Hermitian $(n+1)$-fold with a free $T^2$-action by $g$-sometries. Assume also that $g$ restricted to the $T^2$-fibers is the standard metric on $T^2$. Then $N$ is isomorphic (biholomorphic and isometric) to a manifold $M^{P,Q}$ for some Hermitian $n$-fold $M$ and the closed $2$-forms $\omega_P$ and $\omega_Q$ on $M$ satisfying the conditions of Theorem \ref{main}.
\end{tom1} 
{\bf Proof:}
We have the quotient $M=N/T^2$ and the projection $\pi:N \mapsto M$. Let $H$ be the orthogonal complement to the tangent space to the $T^2$-fibers in $TN$. Thus $H$ is a horizontal distribution (connection) for the principal $T^2$-bundle $N$. 
To construct the line bundles $P$ and $Q$ let $f_1$ and $f_2$ be the
standard 1-dimensional
representations of $T^2$, i.e. \[f_1(e^{ix},e^{iy})(z)=e^{ix}z ~\quad and ~\quad
f_2(e^{ix},e^{iy})(z)=e^{iy}z\] Let $P$ be the line bundle over $M$
associated to $f_1$:
\[P= N \times_{T^2} \mathbb{C}\]
This means that $P$ is the quotient of $N \times \mathbb{C}$ by the
$T^2$-action, where $T^2$ acts
on $\mathbb{C}$ via $f_1$. The construction of $Q$ is similar. Also $H$ induces a connection on both $P$ and $P$. Let $\omega_P$ and $\omega_Q$ be the curvature forms of $P$ and $Q$ correspondingly. We construct $M^{P,Q}$ as in the proof of Theorem \ref{main}. Note that so far we haven't proved that $\omega_P+i\omega_Q$ has no $(0,2)$-component, thus $M^{P,Q}$ in so far just an almost-complex manifold with a Hermitian metric.  

We now construct an isomorphism $\phi: N \mapsto M^{P,Q}$.
It is constructed as follows: take a point $z$ in $N$ and consider $(z,1,1)
\in N \times \mathbb{C} \times \mathbb{C}$.
Taking the orbit of $(z,1,1)$ by the $T^2$-action we get a point $\phi(z)
\in M^{P,Q}$. One easily checks
that $\phi$ is indeed an isomorphism. Hence in particular the almost-complex structure on $M^{P,Q}$ is integrable and by the remark after equation (\ref{dvarphi2}) we conclude that $\omega_P+i\omega_Q$ has no $(0,2)$-component.
Q.E.D.

\section{Intrinsic torsion of $SU(3)$-structures}

In this section we specialize to the case then $dim_{\mathbb{C}}M=2$.
From now on we consider $(M,g_{CY},\Omega^{2,0})$ to be a Calabi-Yau $2$-fold with a non-vanishing holomorphic $(2,0)$-form $\Omega^{2,0}$ which has unit length with respect to $g_{CY}$. If we pick the forms $\omega_P$ and $\omega_Q$ on $M$ satisfying the conditions of Theorem \ref{main} then we obtain a Hermitian $3$-fold $M^{P,Q}$ with a holomorphic $(3,0)$-form $\Omega^{3,0}$ of unit length.
We'll study the intrinsic
torsion of the $SU(3)$-structures we wrote down on $M^{P,Q}$.\\ 
For a general $SU(3)$ structure, Chiossi and
Salamon \cite{CS} have
decomposed this tensor into $5$ components $W_1,\ldots,W_5$. Let
$\Omega^{3,0}_+$ be the real
part of $\Omega^{3,0}$ and let $\Omega^{3,0}_-$ be the imaginary part
of $\Omega^{3,0}$. According to \cite{CS}, there is the following 1-1
correspondence:
\begin{eqnarray}\label{W}
W_1 \longleftrightarrow (d\tilde{\omega})^{3,0}        \nonumber \\
W_2 \longleftrightarrow ((d\Omega^{3,0}_+)_0^{1,1}, (d\Omega^{3,0}_-)_0^{1,1})     \nonumber\\
W_3 \longleftrightarrow (d\tilde{\omega})_0^{2,1}       \nonumber\\
W_4 \longleftrightarrow \tilde{\omega} \wedge d \tilde{\omega}      \nonumber\\
W_5 \longleftrightarrow  (d\Omega^{3,0}_{\pm})^{3,1}  
\end{eqnarray}
The component $W_1$ vanishes because the complex structure on $M^{P,Q}$
is integrable and so the exterior derivative of a $(1,1)$-form has no
$(3,0)$ components.
The components $W_2$ and $W_5$ vanish since $d \Omega^{3,0}=0$. We finally
study the component $W_4$. We need to write down an explicit expression for the
hermitian $(1,1)$-form $\tilde{\omega}$ on $M^{P,Q}$. We
know that $\tilde{\omega}$ equals to $\pi^{\ast}(\omega_{CY})$ on the horizontal
distribution $H$ and its equals to $dx\wedge dy$
on the vertical distribution $V$. In the notation of Theorem \ref{main} we get that
\begin{equation}
\label{omega'}
\tilde{\omega}=\pi^{\ast}\omega_{CY}+ (dx+\pi^{\ast}\alpha) \wedge
(dy+\pi^{\ast}\beta)
\end{equation}
Indeed the 2-form $(dx+\pi^{\ast}\alpha) \wedge (dy+\pi^{\ast}\beta)$
has $H$ as its kernel and it equals to
$dx \wedge dy$ on $V$, hence equation (\ref{omega'}) is true. \\
Let $\omega_P$ be the curvature form of $P$ (so $\omega_P=d\alpha$) and
let $\omega_Q=d\beta$ be the curvature form of $Q$.
We have that
\begin{eqnarray}
\label{domega'}
d\tilde{\omega}= \pi^{\ast}\omega_P \wedge (\pi^{\ast}\beta+dy)-(\pi^{\ast}\alpha+dx) \wedge \pi^{\ast}\omega_Q          
         \nonumber \\
d \tilde{\omega} \wedge \tilde{\omega}= dy \wedge \pi^{\ast}(\omega_P \wedge \omega_{CY})-dx
\wedge \pi^{\ast}(\omega_Q \wedge \omega_{CY})
\end{eqnarray}
The later term vanishes if and only if the $(1,1)$-components of $\omega_P$ and $\omega_Q$ are anti-selfdual. From this we conclude: 

\begin{tom1}
Let $(M,g_{CY},\Omega^{2,0})$ be a Calabi-Yau $2$-fold with a non-vanishing holomorphic $(2,0)$-form $\Omega^{2,0}$. Let $\omega_P$ and $\omega_Q$ be closed $2$-forms on $M$ satisfying the conditions of Theorem \ref{main}. Then the $3$-fold $M^{P,Q}$ has a natural special $SU(3)$-structure with a {\bf constant} dilaton, and the torsional constraints (\ref{torscon}),(\ref{exact}) hold if and only if the $(1,1)$-components of $\omega_P$ and $\omega_Q$ are anti-selfdual: 
$\star\omega^{(1,1)}_P = -\omega^{(1,1)}_P$,
$\star\omega^{(1,1)}_Q = -\omega^{(1,1)}_Q$.
\end{tom1}

\section{Metric scaling and non-constant dilaton}

\label{warp}
We continue to consider $(M,g_{CY},\Omega^{2,0})$ to be a Calabi-Yau $2$-fold with a non-vanishing holomorphic $(2,0)$-form $\Omega^{2,0}$ which has unit length with respect to $g_{CY}$. Furthermore we assume that the $(1,1)$-components
of the forms
$\omega_P$ and $\omega_Q$ are anti-selfdual. In this case $W_4=0=W_5$ and
so the supersymmetry equation
\[2W_4+W_5=0\] certainly holds. We will now define a class $g_{\psi}$ of Hermitian
metrics on $M^{P,Q}$ depending on a function $\psi:M \mapsto \mathbb{R}$
s.t. the supersymmetry equation still holds for them.\\
So let $\psi$ be a function on $M$ and lift it to $M^{P,Q}$. Define the
metric $g_{\psi}$ on $M^{P,Q}$ as in (\ref{metricwarp}). Thus the horizontal and the vertical
distribution
are still perpendicular, on $V$ the metric $g_{\psi}$ is the original metric
$g$ and on $H$ the metric is scaled by $e^{2\psi}$. If we take
$\Omega^{3,0}_{\psi}=e^{2\psi} \Omega^{3,0}$ then it has length 1 with respect to $\tilde{g}_{\psi}$.
We will now show that the supersymmetry equation still
holds for this $SU(3)$-structure: 

\begin{tom1}
Let $(M,\omega_{CY},\Omega^{2,0})$ be a Calabi-Yau $2$-fold with a non-vanishing holomorphic $(2,0)$-form $\Omega^{2,0}$. Let $\omega_P$ and $\omega_Q$ be closed $2$-forms on $M$ satisfying the conditions of Theorem \ref{main}. Assume also that the $(1,1)$-components of $\omega_P$ and $\omega_Q$ are anti-selfdual. Let $\psi: M \mapsto \mathbb{R}$ be any function. Then there is a special $SU(3)$-structure on $M^{P,Q}$ with the Hermitian metric $g_{\psi}$ of the form (\ref{metricwarp}) where $\alpha$ and $\beta$ are as Theorem \ref{main} and the torsional constraints (\ref{torscon}),(\ref{exact}) hold.
\end{tom1}
{\bf Proof:} First we need to explain what $W_4$ and
$W_5$ are.\\
We begin by defining on any Riemannian manifold $N$ a contraction pairing
\[\;\lrcorner\; \, : \: \Lambda^k T^{\ast} N \otimes \Lambda^n T^{\ast} N \mapsto
\Lambda^{n-k} T^{\ast} N \]
(see \cite {lust}, p. 5). This is defined as follows: for an orthonormal
basis $e_1,\ldots,e_l$ of $TN$ let $e^i$ be the dual basis. Let
$\alpha=(\alpha_1,\ldots,\alpha_k)$ be a multi-index of distinct integers
between $1$ and $l=dim(N)$ of length $k$ and let $e^{\alpha}=\bigwedge
e^{\alpha_i}$.
Let $\beta$ be a multi-index of length $n$. We define $e^{\alpha} \;\lrcorner\;
e^{\beta}$ as follows: If the set $(\alpha)$  is not contained in the set
$(\beta)$ then the answer is $0$. If $(\alpha) \subset (\beta)$ then
\[e^{\alpha} \;\lrcorner\; e^{\beta}=(-1)^k e^{\beta-\alpha}\]
Here $(-1)^k$ is the sign of permutation that is needed to put $\alpha$ in
the beginning of $\beta$.
We have the following basic
\begin{prop}
Let $dim(N)=4$ and let $\omega$ be a Hermitian 2-form for the metric on $TN$ (coming from
some compatible almost complex structure). Then for any 1-form $\delta$,
\[\omega \;\lrcorner\; \omega \wedge \delta =\delta\]
\end{prop}
{\bf Proof:} This is linear algebra. We write $\omega=\Sigma dx_i \wedge
dy_i$ at 1 point. It is enough to prove the Proposition for $\delta=dx_i$
or $\delta=dy_j$, and those are immediate. Q.E.D.\\
We now take another proposition from linear algebra that we will need:
\begin{prop}
Let $V$ be a Hermitian vector space of complex dimension $3$ let 
$\Omega$ be a $(3,0)$-from on $V$ of length 1. Then for any 1-form
$\delta$ on $V$ we'll have that \[Re\; \Omega \;\lrcorner\; \delta \wedge Re\;
\Omega = -2 \delta\]
\end{prop}
{\bf Proof:} Obviously it is enough to prove it for a real 1-form
$\delta$.  We certainly have a subspace $W$ of $V$ of complex dimension
$1$ on which $\delta$ vanishes. Let $U=W^{\perp}$, then $\delta$ can be
viewed as a form on $U$. Let $\xi$ be a $(2,0)$-form on $U$ of length $1$.
We can choose orthonormal basis $e_1$ and $e_2=Je_1$ on $W$
such that $\Omega=\xi \wedge (e^1+ie^2)$. Then $Re\; \Omega= Re\; \xi
\wedge e^1-Im\; \xi \wedge e^2$. Here $Re\; \xi$ and $Im\; \xi$ are hermitian
2-forms for the metric on $U$ (for different complex structures on $U$). In
particular the previous proposition holds for them.\\
We have that $\delta \wedge Re\; \Omega=-(e^1 \wedge Re\; \xi \wedge
\delta-e^2 \wedge Im\; \xi \wedge \delta)$. From this we immediately derive
the statement of our proposition. Q.E.D.\\
\\
We now return to $M^{P,Q}$ with the metric $\tilde{g}_{\psi}$.
We have a form $\Omega^{3,0}_{\psi}=e^{2\psi}\Omega^{3,0}$ and a Hermitian form
\[\tilde{\omega}_{\psi}=\rho \wedge \bar{\rho}+e^{2\psi} \pi^{\ast}\omega_{CY}\]
The classes $W_4$ and  $W_5$ can be written in a form:
\[W_4={1\over 2}\,  \tilde{\omega}_{\psi} \;\lrcorner\; d \tilde{\omega}_{\psi} ~ , ~ W_5={1\over 2}\,  Re\;
\Omega^{3,0}_{\psi} \;\lrcorner\; d Re\;\Omega^{3,0}_{\psi}\]
(see \cite{lust}, p. 5). We compute that \[d\tilde{\omega}_{\psi}=e^{2\psi}2d\psi\wedge
\pi^{\ast}\omega_{CY}-\pi^{\ast}\omega_P \wedge (dy+\beta)+\pi^{\ast}\omega_Q
\wedge (dx+\alpha)\]
Note that \[\;\lrcorner\;: \Lambda^2T^{\ast} \otimes \Lambda^2 T^{\ast}
\mapsto \mathbb{R}\] is just the dot product. Since the dot product of
$\omega_{CY}$ with $\omega_P$ and $\omega_Q$ is $0$, we conclude that
\[\tilde{\omega}_{\psi} \;\lrcorner\; (-\pi^{\ast}\omega_P \wedge (dy+\beta)+\pi^{\ast}\omega_Q
\wedge (dx+\alpha))=0\]
Also \[\tilde{\omega}_{\psi} \;\lrcorner\; e^{2\psi}2d\psi \wedge \pi^{\ast}\omega_{CY}=\tilde{\omega}_{\psi} \;\lrcorner\; 2d\psi \wedge e^{2\psi}\pi^{\ast}\omega_{CY} =2d\psi\]
Also $d\,Re\; \Omega^{3,0}_{\psi}=2d\psi \wedge Re\; \Omega^{3,0}_{\psi}$. Hence \[ Re\; \Omega^{3,0}_{\psi}
\;\lrcorner\; 2d\psi \wedge Re\; \Omega^{3,0}_{\psi} =-4d\psi\]
From all this we conclude that $2W_4+W_5=0$. Q.E.D.

\section{Examples}\label{K3}

In this section, we will consider examples when Theorem \ref{dilaton} applies. 
\subsection{$SU(3)$-fibrations over K3 surfaces}
Let $M$ be a K3 surface with
a Calabi-Yau metric $g_{CY}$ and a holomorphic $(2,0)$-form $\Omega^{2,0}$. Let
$C_1,\ldots,C_k$ be some collection of holomorphic curves
on $M$ (e.g. if $M$ is a Kummer $K3$ and $C_i$ are the exceptional spheres).
We consider the divisor \[C=\sum a_k C_k\]
Here $a_k$ are integers such that \[\sum a_i \int_{C_i} \omega_{CY}=0\]
We will only consider the case when such $a_k$ exist, {\it e.g.} a Kummer $K3$.
$C$ defines a lines bundle $P$ with a meromorphic section $\sigma$ and the
first Chern class $c_1(P)$ is the Poincare dual of $C$, hence
it satisfies
\begin{equation}
\label{c1}
c_1(P) \wedge [\omega_{CY}]=0 ~ , ~ c_1(P) \wedge [Re\; \Omega^{2,0}]=0 ~ , ~ c_1(P)
\wedge [Im\; \Omega^{2,0}]= 0
\end{equation}
We now use the fact that $b_2^{+}(M)=3$ and $[\omega_{CY}] ~ , ~ [Re\; \Omega^{2,0}]$ and $ [Im\; \Omega^{2,0}]$ is a basis for $H^2_+(M)$. Let $\frac{-\omega_P}{2\pi} $ be the
harmonic representative of $c_1(P)$.
Equation (\ref{c1}) implies that $\omega_P$ is anti-selfdual, hence
in particular it is of type $(1,1)$.
We also choose $\omega_Q$ by the same principle as $\omega_P$. Theorem \ref{dilaton} applies.
\subsection{$SU(3)$-fibrations over $T^4$}
We choose a standard flat metric on the four-dimensional torus. We have the following basis of antiselfdual $2$-forms on $T^2$:
\[\omega_1=2\pi(dx_1\wedge dx_2-dx_3\wedge dx_4)\]
\[\omega_2=2\pi(dx_1\wedge dx_3-dx_4\wedge dx_2)\]
\[\omega_2=2\pi(dx_1\wedge dx_4-dx_2\wedge dx_3)\]
We can choose $\omega_P$ and $\omega_Q$ to be linear combinations of $\omega_1,\omega_2,\omega_3$ with integer coefficients and Theorem \ref{dilaton} applies.
\section{Cohomology and Hodge numbers of $M^{P,Q}$}

In this section, we study the cohomology and the Hodge numbers $h^{1,0}$ and $h^{0,1}$ of $M^{P,Q}$. For Hodge numbers  we assume that $M$ is compact and that $\omega_P$ and $\omega_Q$ are of type $(1,1)$.

\subsection{Hodge numbers $h^{1,0}$ and $h^{0,1}$ of $M^{P,Q}$}

We note that if $\xi$ is any harmonic form (both for the usual
Laplacian or for the $\overline{\partial}$-Laplacian) then the $\partial_x$
and the $\partial_y$-flows are structure preserving and they preserve the
cohomology class of $\xi$, hence they preserve $\xi$. This is clear for the usual cohomology. For the  $\overline{\partial}$-cohomology we note that the $\partial_x$-flow acts upon the space $H^{p,q}$. So we have a one dimensional representation of the circle on $H^{p,q}$ with weights $iA_1,\ldots,iA_k$. Here $A_i$ are integers. Also the $\partial_y$-flow acts on $H^{p,q}$. For any harmonic form $\xi$ in $H^{p,q}$ we have that $L_{\partial_x}\xi+iL_{\partial_y}\xi$ is in $H^{p,q}$. But the $\Lambda^{p,q}$-component of $L_{\partial_x}\xi+iL_{\partial_y}\xi$ is $\overline{\partial}(i_{\partial_x+i\partial_y}\xi)$. 
From this we deduce that $L_{\partial_x}\xi+iL_{\partial_y}\xi=0$. This implies that the representation of $\partial_y$ on $H^{p,q}$ has weights $A_1,\ldots,A_k$. But $\partial_y$ is periodic, hence $A_i=0$, i.e. the $\partial_x$ and the $\partial_y$-flow preserves $H^{p,q}$.

 Let $\xi$ be a harmonic $(1,0)$-form on $M^{P,Q}$. The invariance of $\xi$ under the $\partial_x$
and the $\partial_y$-flow implies that
one can write \[\xi=A \rho+ \pi^{\ast}s_1\]
Here $s_1$ is a $(1,0)$-form on $M$ and $A$ is a function pulled up from $M$. Since $\overline{\partial}\rho=\pi^{\ast}(\omega_P+i\omega_Q)$, the equation $\overline{\partial}\xi=0$
translates into $\overline{\partial}A=0$ and $A(\omega_P+i\omega_Q)+\overline{\partial}\phi=0$. So $A$ is a constant and moreover $\omega_P+i\omega_Q$ is a non-zero harmonic $(1,1)$-form for the $\overline{\partial}$-complex. Hence $A=0$ and $ \overline{\partial}\phi=0$, i.e.  $H^{1,0}(M^{P,Q}) \simeq H^{1,0}(M)$. In the $H^{0,1}$-case simple analysis shows that harmonic $(0,1)$ forms are all of the form $c\bar{\rho}+\pi^{\ast}s_1$ for $s^1 \in H^{0,1} M$ and $c$ a constant. Thus $h^{0,1}(M^{P,Q})=h^{0,1}(M)+1$. In particular $h^{0,1}(M^{P,Q})=h^{1,0}(M^{P,Q})+1$.

\subsection{Cohomology of $M^{P,Q}$}

To study cohomology first let $M^{P}$ be the unit circle bundle of $P$. Certainly $M^P$ is a circle bundle over $M$. Also $M^{P,Q}$ is a circle bundle over $M^P$ (with the fiber being the unit circle of $Q$ pulled up to $M^P$). We'll use Gysin sequence (see \cite{Hus}, p. 255) to study the cohomology one step at a time. The Gysin sequence tells that if $F$ is a circle bundle with projection $\pi$ over the base $B$ and the first Chern class of $F$ is $c_1 \in H^2(B,\mathbb{R})$ then we have the following exact sequence:
\[H^i B \stackrel{\cup c_1}{\longrightarrow}H^{i+2}B \stackrel{\pi^{\ast}}{\longrightarrow} H^{i+2}F \rightarrow H^{i+1}B \rightarrow \ldots\] 
Let's study $H^1$ first. Since $\omega_P$ is non-trivial in real cohomology we conclude that $H^1M^P \simeq H^1 M$. Now we have 2 cases:\\
{\bf Case 1:} If $\omega_Q$ is not a multiple of $\omega_P$ in $H^2M$ then the Gysin sequence tells that it lifts to a non-trivial element of $H^2 M^P$. Using the Gysin sequence again for the fibration $M^{P,Q} \mapsto M^P$ we conclude that $H^1 M^{P,Q} \simeq H^1 M^P \simeq H^1 M$.\\
{\bf Case 2:} If $\omega_Q$ is a multiple of $\omega_P$ in $H^2$ then Gysin sequence tells that it lifts to a trivial class in $H^2 M^P$. Using the Gysin sequence again for the fibration $M^{P,Q} \mapsto M^P$ we conclude that $b_1(M^{P,Q})=b_1(M^P)+1=b_1(M)+1$.\\
{\bf Remark:} The Gysin sequence implies that all the first cohomology of $M^P$ comes from $M$ and so the circle fibers are trivial in the first homology of $M^P$. From this we conclude that the torus fibers are trivial in the second homology of $M^{P,Q}$, we used this fact before to show that $M^{P,Q}$ admits no K\"ahler metric.\\
Let us consider $H^2$. We assume that $H^1(M)=0$. Using the Gysin sequence we conclude that $b_2(M^P)=b_2(M)-1$. Now we have 2 cases:\\
{\bf Case 1:} If $\omega_Q$ is not a multiple of $\omega_P$ in $H^2M$ then the Gysin sequence tells that it lifts to a non-trivial element of $H^2 M^P$. Using the Gysin sequence again for the fibration $M^{P,Q} \mapsto M^P$ we conclude that $b_2(M^{P,Q})=b_2(M^P)-1=b_2(M)-2$.\\
{\bf Case 2:} If $\omega_Q$ is a multiple of $\omega_P$ in $H^2$ then Gysin sequence tells that it lifts to a trivial class in $H^2 M^P$. Using the Gysin sequence again for the fibration $M^{P,Q} \mapsto M^P$ we conclude that  $b_2(M^{P,Q})=b_2(M^P)=b_2(M)-1$.

This information is enough to find all the Betti numbers in case $M$ is a $K3$-surface. The only unknown so far is $b_3(M^{P,Q})$ and we can find it using the fact that the Euler characteristic of $M^{P,Q}$ is $0$ (because $\partial_x$ is a non-vanishing vector field on $M^{P,Q}$).

\section{Pulling up Special Lagrangians}

Here we assume that $M$ is a Calabi-Yau manifold and $\omega_Q=0$. We take $Q$ to be the trivial bundle with the trivial connection. Thus $M^{P,Q}$ is a direct product of the unit
circle bundle $M^P$ on $P$ with the unit circle $S^1$: \[M^{P,\mathbb{C}} 
\simeq M^P \times
 S^1\] For any element $s \in S^1$ let \[M^{P,s}= M^P \times (s)\] The
horizontal distribution $H$ along $M^{P,s}$ is tangent to $M^{P,s}$. Moreover 
there is a natural circle action on $M^{P,s}$ given by
\[e^{ix}(\xi,s)=(e^{ix}\xi,s)\]
The  vector field generating this action is $\partial_x$.\\
If $L$ is a submanifold of $M$ then we can define its lift
\[L_s=\pi^{-1}(L) \bigcap M^{P,s}\] We have that the tangent space to $L_0$ 
naturally splits as
\[TL^ H \oplus span(\partial_x)\] Here $TL^H$ is the horizontal lift of $TL$ to
$H$. From this we get the following:
\begin{prop}
Let $L$ be a Special Lagrangian submanifold on $M$. Then $L_s$ is a
Special Lagrangian submanifold of $M^{P,\mathbb{C}}$.
\end{prop}
We wish to point out that if we have a Special Lagrangian fibration on $M$
it lifts to a special Lagrangian fibration on $M^{P,\mathbb{C}}$.

\section*{Conclusion}

In this paper, we presented a geometric construction for complex non-K\"ahler manifolds with intrinsic 
$SU(3)$ structure, used in supersymmetric string compactifications.\footnote{It should be mentioned
that any realistic heterotic string compactification model has to include a vector bundle construction,
with a gauge group rich enough to account for the Standard Model particles. This requirement can impose
severe constraints on the geometric construction, {\it e.g.}, one might need to consider singular limits
of the base CY 2-fold. We do not study this subject here.} 
We gave
a general construction of families $M^{P,Q}$ of holomorphic $T^2$-fibrations over a Hermitian $n$-fold $M$.
We have shown that our construction can satisfy the supersymmetry conditions in the heterotic string theory,
and in this case we get complex half-flat, or special Hermitian manifolds. We presented examples 
of $T^2$- bundles over $K3$ surfaces and a four-torus which satisfy the supersymmetry constraint.

Also, we proposed a modified model with the metric depending on a warp factor, and shown that the supersymmetry conditions are still satisfied. In addition, we computed all Betti numbers and
the Hodge numbers $h^{1,0}$ and $h^{0,1}$. 

It is also shown that in the case $\omega_Q=0$ one can lift Special Lagrangian submanifolds and
fibrations from the Calabi-Yaus to the bundles $M^{P,Q}$, and Special 
Lagrangian submanifolds on those bundles are still calibrated by $Re\;  \Omega^{3,0}$ (and hence minimal).

\bigskip 
\noindent
{\bf Acknowledgments}
\medskip

We would like to acknowledge very useful conversations with
K.~Dasgupta, S. Gukov, S.~Kachru, and L.~McAllister. 

The work of S.~P. was supported by Stanford Graduate Fellowship.


\begin{thebibliography}{4}

\bibitem{bd}
K.~Becker, K.~Dasgupta, ``Heterotic Strings with Torsion'', {\tt hep-th/0209077}.

\bibitem{kstt}
S.~Kachru, M.B.~Schulz, P.~K.~Tripathy, and S.~P.~Trivedi,
 ``New Supersymmetric String Compactifications,''  {\tt hep-th/0211182}.

\bibitem{louis}
 S.~Gurrieri, J.~Louis, A.~Micu, D.~Waldram, ``Mirror Symmetry in Generalized Calabi-Yau Compactifications'',
{\tt hep-th/0211102}.

\bibitem{lust}
 G.~L.~Cardoso, G.~Curio, G.~Dall'Agata, D.~Luest, P.~Manousselis, G.~Zoupanos, 
``Non-Kaehler String Backgrounds and their Five Torsion Classes'',  {\tt hep-th/0211118}.

\bibitem{gm}
 S.~Gurrieri, A.~Micu,
``Type IIB Theory on Half-flat Manifolds", {\tt hep-th/0212278}.

\bibitem{bbdg}
K.~Becker, M.~Becker, K.~Dasgupta, P.~Green,  ``Non-K\"ahler Compactifications",  in preparation.

\bibitem{tt}
P.~K.~ Tripathy and S.~P.~Trivedi, 
``Compactifications with Flux on $K3$ and Tori,"  in preparation.


\bibitem{fg} Anna Fino, Gueo Grantcharov ``On some properties of the manifolds with skew-symmetric torsion and holonomy SU(n) and Sp(n)'', math.DG/0302358

\bibitem{Gaud} Gauduchon, Paul ``Hermitian connections and Dirac operators.''
Boll. Un. Mat. Ital. B (7) 11 (1997), no. 2, suppl., 257--288. 

\bibitem{Str}
A.~Strominger, ``Superstrings with torsion,'' {\em Nucl. Phys.} {\bf B274}
  (1986) 253.

\bibitem{kst}
S.~Kachru, M.~Schulz, and S.~Trivedi, ``Moduli Stabilization
from Fluxes in a Simple IIB Orientifold,'' {\tt hep-th/0201028}.

\bibitem{SS}
J.~Scherk and J.~H.~Schwarz,  ``How To Get Masses From Extra Dimensions,''
{\em Nucl.\ Phys.\ B} {\bf 153}, 61 (1979).

\bibitem{Hull}
C.~M.~Hull, ``Superstring Compactifications With Torsion And
Space-Time Supersymmetry,'' in {\it Turin 1985, Proceedings,
Superunification and Extra Dimensions}, 347-375.

\bibitem{FFS}
M.~Falcitelli, A.~Farinola, and S.~Salamon,
``Almost-Hermitian Geometry,"
{\em Diff.\ Geo.\ } {\bf 4} (1994) 259.

\bibitem{KM}
N.~Kaloper and R.~C.~Myers,
``The O(dd) story of massive supergravity,''
JHEP {\bf 9905}, 010 (1999), {\tt hep-th/9901045}.

\bibitem{Vafa}
C.~Vafa, ``Superstrings and topological strings at large N'',
{\em J.\ Math.\ Phys.\ } {\bf 42} (2001) 2798,
{\tt hep-th/0008142}.

\bibitem{joyce}
D.~Joyce, ``Compact Manifolds with Special Holonomy", Oxford University Press,
Oxford, 2000.

\bibitem{fi}
T.~Friedrich and S.~Ivanov, "Parallel spinors and connections with
skew-symmetric torsion in string  theory," {\tt math.dg/0102142}.

\bibitem{salamon1}
S.~Salamon, {\it Complex structures on nilpotent {Lie} algebras},  {\em J Pure
  Appl Algebra} (1998) {\tt math.DG/9808025}.

\bibitem{drs}
K.~Dasgupta, G.~Rajesh, S.~Sethi,
``M Theory, Orientifolds and G-Flux",   {\em JHEP} {\bf 9908} (1999) 023, {\tt hep-th/9908088}.

\bibitem{salamon}
S.\ Salamon, ``Almost Parallel Structures'', 
in \emph{Global Differential Geometry: The Mathematical Legacy of Alfred Gray
(Bilbao, 2000)}, pp. 162, {\tt math.DG/0107146}.

\bibitem{CS}
S.\ Chiossi and S.\ Salamon, ``The Intrinsic Torsion of $SU(3)$ and $G_2$
Structures,'' in \emph{Differential geometry, Valencia, 2001}, pp. 115, 
{\tt math.DG/0202282}. 

\bibitem{rocek}
M.~Rocek,
``Modified Calabi--Yau manifolds with torsion,''
in {\it Essays on Mirror Manifolds}, ed.\ S.T.\ Yau, International Press,
Hong Kong, 1992;\\
S.~J.~Gates, C.~M.~Hull, and M.~Rocek,
``Twisted Multiplets And New Supersymmetric Nonlinear Sigma Models,''
{\em Nucl.\ Phys.\ } {\bf B248} (1984) 157;\\
S.~Lyakhovich and M.~Zabzine,
``Poisson geometry of sigma models with extended supersymmetry,''
{\em Phys.\ Lett.\ B} {\bf 548} (2002) 243
{\tt hep-th/0210043}.

\bibitem{ip}
S.~Ivanov and G.~Papadopoulos,
``Vanishing theorems and string backgrounds,''
{\em Class.\ Quant.\ Grav.\ } {\bf 18} (2001) 1089,
{\tt math.dg/0010038};\\
``A no-go theorem for string warped compactifications,''
{\em Phys.\ Lett.\ } {\bf B497} (2001) 309, {\tt hep-th/0008232};\\
G.~Papadopoulos,
``KT and HKT geometries in strings and in black hole moduli spaces,''
{\tt hep-th/0201111};\\
J.~Gutowski, S.~Ivanov, and G.~Papadopoulos,
``Deformations of generalized calibrations and compact non-K\"ahler manifolds 
with vanishing first Chern class,''
{\tt math.dg/0205012}.

\bibitem{gkmw}
J.~P.~Gauntlett, N.W.~Kim, D.~Martelli, and D.~Waldram,
``Fivebranes wrapped on SLAG three-cycles and related geometry,''
{\em JHEP} {\bf 0111} (2001) 018, {\tt hep-th/0110034};\\
J.~P.~Gauntlett, D.~Martelli, S.~Pakis, and D.~Waldram,
``G-structures and wrapped NS5-branes,'' {\tt hep-th/0205050}.

\bibitem{kmpt}
P.~Kaste, R.~Minasian, M.~Petrini, and A.~Tomasiello,
``Kaluza-Klein bundles and manifolds of exceptional holonomy,"
{\em JHEP} {\bf 0209} (2002) 033,
{\tt hep-th/0206213}.

\bibitem{hmw}
S.~Hellerman, J.~McGreevy, B.~Williams,
``Geometric Constructions of Nongeometric String Theories", {\tt hep-th/0208174}.

\bibitem{Cal}
E.~Calabi ``A class of compact, complex manifolds which are
   not algebraic.'' Ann. of Math. (2) 58, (1953). 494--500.

\bibitem{Hitchin}
N.~Hitchin, ``Stable forms and special metrics,''  in {\em Global
  Differential Geometry: The mathematical legacy of Alfred Gray}, pp.~70--89,   AMS, 2001.

\bibitem{FPS} A.~Fino, M.~Parton, S.~Salamon: ``Families of strong KT structures in six dimensions,'' 
{\tt math.DG/0209259}.

\bibitem{GH} P.~Griffiths, J.~Harris, ``Principles of Algebraic geometry,'' Wiley and Sons,1978.

\bibitem{Hus} D.~Husemoller, ``Fiber bundles'', Springer Graduate texts in mathematics, 3rd Edition.


\end{thebibliography}
\end{document}